\documentclass[12pt, superscriptaddress]{revtex4-1}

\usepackage{graphicx}
\usepackage{subfigure}
\usepackage{epstopdf}

\begin{document}

\title{Activity report of ILD-TPC Asia group}
\noaffiliation
\author{Y.~Kato}
\thanks{e-mail address: katoy@hep.kindai.ac.jp \\
Talk presented at the International Workshop on Future Linear Colliders (LCWS13), Tokyo, Japan, 11-15 November 2013.}
\affiliation{Kinki U., 3-4-1 Kowakae, Higashi-osaka 577-8502, Japan}
\author{R.~Yonamine}\affiliation{High Energy Accelerator Research Organization (KEK), 1-1 Oho, Tsukuba 305-0801, Japan}
\author{P.~Gros}\affiliation{Saga U., 1 Honjomachi, Saga 840-8502, Japan}
\author{J.~Tian}\affiliation{High Energy Accelerator Research Organization (KEK), 1-1 Oho, Tsukuba 305-0801, Japan}
\author{S.~Kawada}\affiliation{Hiroshima U., 1-3-1 Kagamiyama, Higashi-Hiroshima 739-8530,  Japan}
\author{K.~Fujii}\affiliation{High Energy Accelerator Research Organization (KEK), 1-1 Oho, Tsukuba 305-0801, Japan}
\author{T.~Matsuda}\affiliation{High Energy Accelerator Research Organization (KEK), 1-1 Oho, Tsukuba 305-0801, Japan}
\author{A.~Sugiyama}\affiliation{Saga U., 1 Honjomachi, Saga 840-8502, Japan}
\author{O.~Nitoh}\affiliation{Tokyo U. of Agriculture and Technology, 2-24-16 Nakamachi, Koganei 184-8588, Japan}
\author{T.~Watanabe}\affiliation{Kogakuin U., 1-24-2 Nishi-Shinjuku, Shinjuku 163-8577 , Japan}
\author{T.~Fusayasu}\affiliation{Nagasaki Institute of Applied Science (NiAS), 536 Abamachi, Nagasaki 851-0193, Japan}
\author{T.~Takahashi}\affiliation{Hiroshima U., 1-3-1 Kagamiyama, Higashi-Hiroshima 739-8530,  Japan}
\author{M.~Kobayashi}\affiliation{High Energy Accelerator Research Organization (KEK), 1-1 Oho, Tsukuba 305-0801, Japan}

\begin{abstract}
The purpose of ILD-TPC Asia group is realization of high precision 
Time Projection Chamber (TPC) with 
Gas Electron Multiplier (GEM) as a central tracker in 
International Linear Collider (ILC). We have been studying the many
R\&D items to build the real detector as a member of LCTPC collaboration.
This paper describes the our efforts for realization of the ILD-TPC,
the result of test beam using large prototype TPC, local field distortion, 
positive ion effects and gate devices, and cooling electronics which are 
key items to build ILD-TPC.
\end{abstract}

\maketitle

\section{Introduction}
The International Linear Collider (ILC) has possibility of 
the discovery of the beyond standard model~\cite{ILC-DBD}. ILC is proposed 
to build two detector systems, ILD~\cite{ILC-ILD} and SiD. The ILD concept
is optimized for particle flow analysis(PFA)~\cite{PFA} aiming at
measuring every particle in an event with higher precision. 
PFA requires the combination of three detectors, trackers, 
electromagnetic calorimeters (ECAL), and hadronic calorimeters (HCAL) 
to reconstruct the every particles, even inside jets. 
The requirement of the central tracker of ILC are; (1) momentum 
resolution is less than $2 \times 10^{-5}$ GeV$^{-1}$ and 
(2) good 2-hit resolution ability.

The central tracker of the ILD is a Time Projection Chamber (TPC) 
which measures the charged track with a large number of 
three-dimensional ($r,\phi, z$) space points and has a minimum
amount of material budget. The size of ILD-TPC is 329~mm for inner radius,
1808~mm for outer radius, and 2350 $\times$ 2~mm (divides two parts) for $z$\
direction~\cite{ILC-DBD}. And ILD-TPC is inside of magnet with 3.5~T and 
has 224 measuring
layers. Under these 
conditions, ILD-TPC has to have the spatial resolution with less than
100~${\rm \mu m}$\ and 2-hit resolution with less than 2~mm 
to satisfy the above requirement (1) and (2).
The traditional TPC has a read out devices by multi-wire proportional
chambers (MWPC). But under the ILD condition which is the magnetic field
of $B = 3.5$\ T, the strong $E \times B$\ effect leads to the deterioration
of the spatial resolution. Therefore other read out devices are needed for
the TPC read out system.

LCTPC collaboration~\cite{ILC-TPC Colla} formed to join the
international efforts to develop ILD-TPC. The group from
Asia, Europe, and America are involved in this research.
LCTPC collaboration has investigated the use of Micro Pattern 
Gaseous Detector (MPGD) as the read out system. The characteristic
of MPGD is the gas amplification in the small pitch of order 
10 - 100 $\mu$m. This improves the spatial 
resolution and reduce of
$E \times B$\ effect.
 
ILD-TPC Asia group~\cite{ILD-TPC Asia} is consist of Japan (KEK, Kinki U., Saga U., 
Tokyo U.of A\& T, Kogakuin U., Hiroshima U., NiAS), China (Tsinghua U.),
and Philippines(Mindanao U.) which join LCTPC collaboration.
The aims of ILD-TPC Asia group is development of TPC with GEM (Gas
Electron Multiplier)~\cite{GEM}
appropriate for ILD-TPC. GEM is a foil made of a metal-insulator-metal
sandwich and has many microscopic holes with the microscopic pitch.
GEM used by our group consists of a 5~$\mu$m copper layer on both side
of a 100~$\mu$m LCP layer. And it has a keystone like shape of dimension 17 $\times$\ 22 cm$^2$.
 The holes  are drilled by the laser with a diameter
 of 70~$\mu$m and make the hexagonal pattern of a pitch of 140~$\mu$m.
When the electron released in  the gas pass through the GEM holes, the electric 
potential difference between the upper and lower copper side amplifies the
electrons. The double GEM layer system has been introduced to obtain the sufficient 
electron amplification. 

\section{Test beam and Analytic formula}
\begin{figure}[htbp]
\begin{center}
\includegraphics[width=8cm, bb=0 0 960 720]{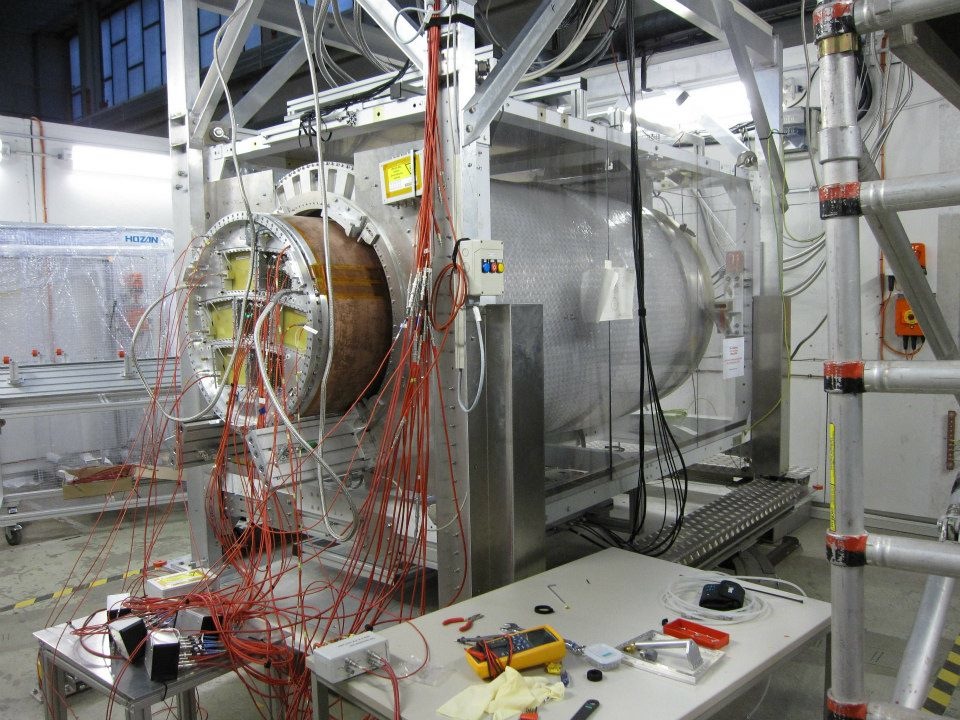}
\caption{TPC LP system in the DESY II test beam area T24/1}
\label{fig:LP-pic} 
 \end{center}
 \end{figure}

LCTPC collaboration has built the TPC large prototype (LP) system in the DESY II test beam area
T24/1~\cite{largeprotype}.  The purpose of this LP is pushing forward the substantial study of MPGD TPC in the situation
that is near to an actual conditions. 
FIG.~\ref{fig:LP-pic} is a picture of LP system. 
LP consists of a large cylindrical field cage with a modular 
end plate, allowing up to 7 detector modules to be mounted. LP is inserted in the PCMAG
superconducting magnet of $B = 1$T.  The size of the field cage is 72 cm $\times$\ 61 cm.
We built 3 double GEM layer modules and  have taken four times of the test beam with LP from 2009 to 2012 using $p = 5$ GeV/c electron beam.
This section explains the result of 2010 and 2012 test beam. 

The 2010 test beam didn't have enough data  because the discharge which was 
occurred on GEM damaged 
the read out electronics. And wrong  voltage setting of the field shaper caused  a large 
distortion around the module boundary (See FIG.~\ref{fig:distortion}). Based on failure of 2010 test beam, there were some 
improvement for 2012 test beam, (1) optimization of the field shaper's voltage setting,
(2) increase the partitions (2 to 4) of GEM electrode to protect the read out electronics by
discharge, (3) add protection circuit to read out electronics. These improvement led to
get the many various data (with and without magnetic field, scanning 
$x$-direction,
scanning $\theta$\ and $\phi$\ direction, etc) on 2012 test beam.

\begin{figure}[htbp]
\centering
\includegraphics[width=9cm, bb= 0 0 729 436]{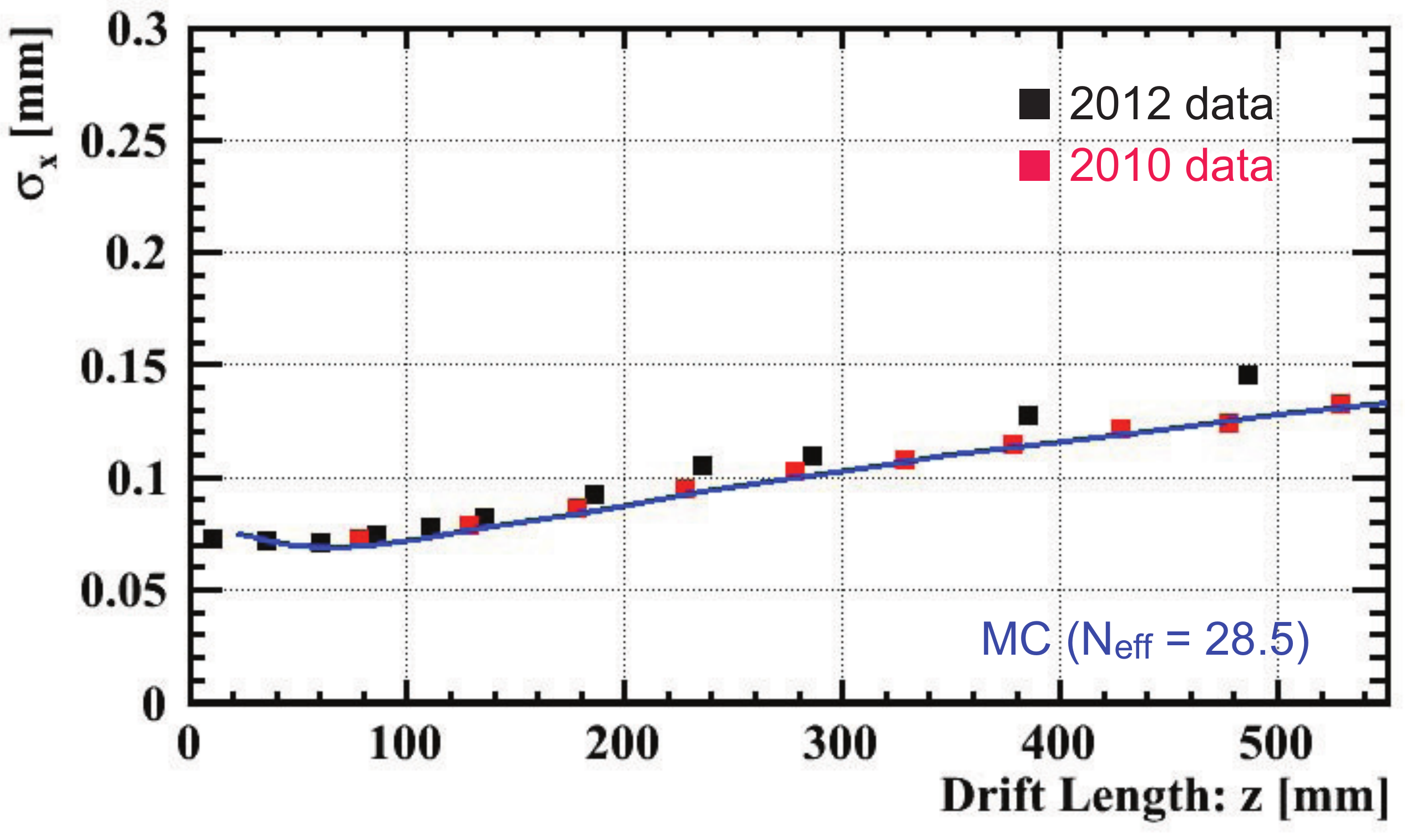}
\caption{Dependence of the drift length on the transverse spatial  resolution}
\label{fig:testbeamresult} 
 \end{figure}

\begin{figure}[htbp]
\centering
\includegraphics[width=11cm, bb=0 0 567 386]{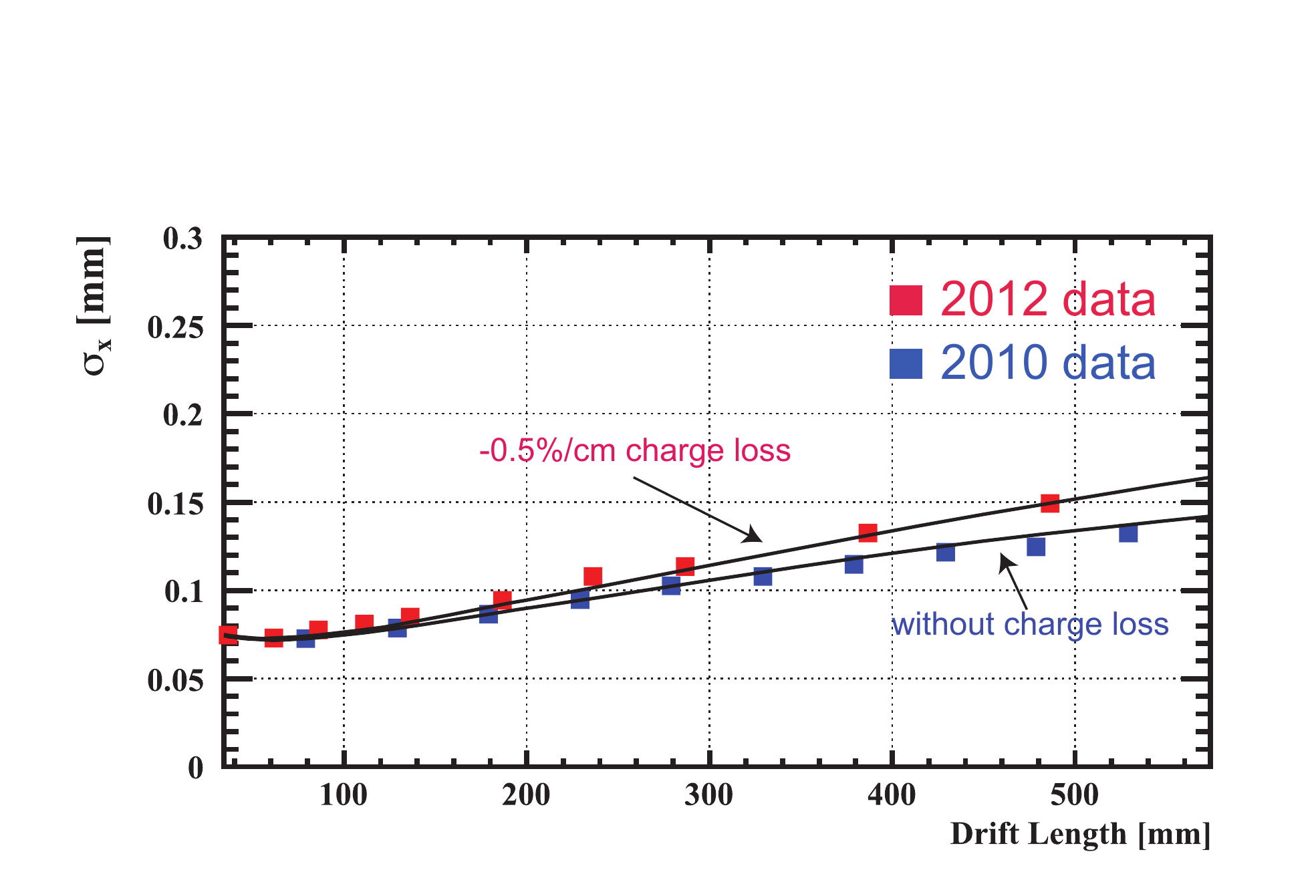}
\caption{Comparison test beam results and simulation}
\label{fig:compdat}
\end{figure}

FIG.~\ref{fig:testbeamresult} shows the dependence of the drift length on the transverse point resolution.
In this figure, there is a different behavior between 2010 and 2012 data. The resolution on
2012 data turns worse as the drift length become long rather than 2010 data.
We tried to find the cause of this deterioration, but  it  can't be explained by  gain uniformity,  discharge effect, cross talk, threshold effect, etc~\cite{JumpingECFA}.
This figure also shows that 2012 data has some data points nearer than 
10~cm and these data seems consistent with 2010 data within 10~cm. 
We conclude the deterioration of the resolution is caused by a charge loss somehow 
because the influence of the charge loss affects longer drift length. 
The charge loss may be happened by gas leak or not correct value of the 
oxygen monitor. 2010 and 2012 data are compared with Monte Carlo 
simulation by our analytic formula~\cite{yonamine}  which is 
based on 2010 data ($N_{\rm eff} = 28.5$). 
FIG.~\ref{fig:compdat} shows the comparison result.
The figure  shows 2010 data consists with the simulation without charge loss and
2012 data consists with the simulation with -5\%/cm charge loss. And the simulation
shows the influence of charge loss is small at the short drift length.
The charge loss by the really lost seed electron can explain 
the change for the deterioration of the transverse resolution on 2012 data.
As the conclusion, 
2012 data is consistent with 2010 data taking account into the real charge loss. 
 
\begin{figure}[htbp]
\centering
 \includegraphics[width=10cm, bb=0 0 233 133]{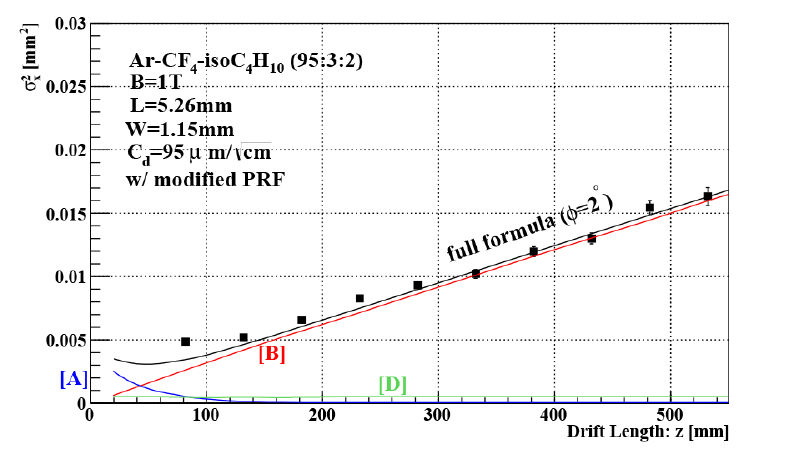}
\caption{Spatial resolution as a function of drift length. The plot shows
${\sigma_{s}}^{2}$\ of the 2010 data together with analytic formula. 
Black circle is 2010 data. [A] is the systematic term, [B] is the diffusion
term and [D] is the cluster term.}
\label{fig:formula}
 \end{figure}

\begin{figure}[htbp]
\begin{center}
 \includegraphics[width=12.5cm, bb=0 0 567 260]{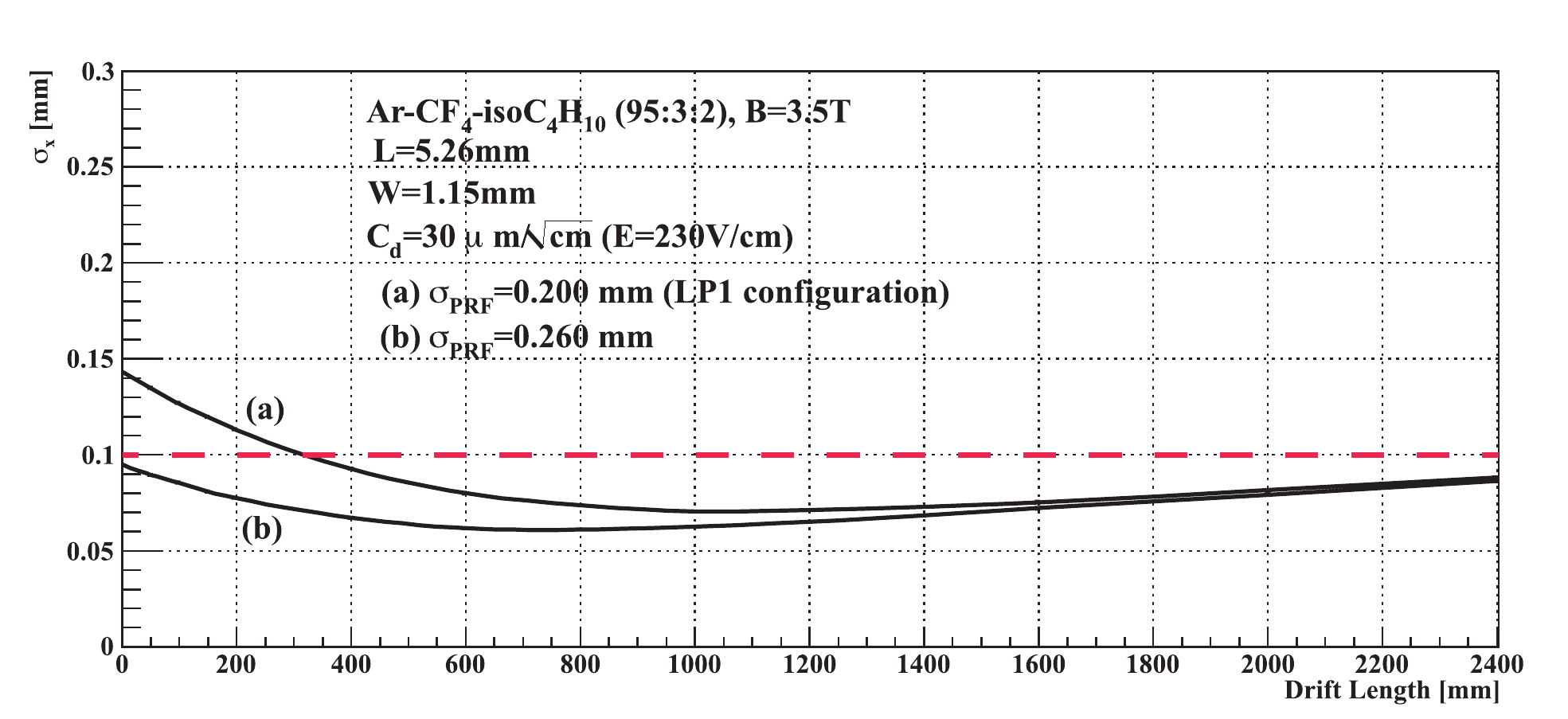}
\caption{The expect performance  of the spatial resolution for ILD-TPC by 2010 test beam. (a) is estimated by Magboltz for B=3.5 T with the same configuration
as our test beam. (b) shows a virtual case of $\sigma_{PRF} = 0.260$\ mm.}
\label{fig:extrpol}
 \end{center}
 \end{figure}

 In order to estimate the performance of real ILD-TPC with the result of the test beam data,
 it is important to  establish the principle formula for the spatial resolution. 
 Our group has developed a new analytic formula which is applicable to titled 
 tracks~\cite{yonamine} . The special resolution consists of four components, 
 [A] systematics due to finite pad read out, [B] diffusion effect, 
[C] electron noise
 effect, and [D] primary cluster fluctuation. 
 Term [D] makes effort to resolution of 
 inclined track. By examining physical meanings of each term of the formula,
 we can identify what factor mainly contributes to the spatial resolution, which help us find
 a way to improve the spatial resolution. This formula attempts to 2010 data.
 FIG.~\ref{fig:formula} shows the result of comparison 2010 data and the formula.
 The line of the formula is not include the term [C] assumed to be negligible.
FIG.~\ref{fig:formula} shows the formula to be consistent with 2010 data and 
the formula shows the track  to be inclined $\phi = 2^{\circ}$. Using this formula and the
the basic parameter for the spatial resolution such as, $C_{\rm d}, PRF, 
N_{\rm eff}$\
and $\hat{N}_{\rm eff}$ obtained by the 2010 data, we can estimate the performance of 
the ILD-TPC. FIG.~\ref{fig:extrpol} shows the expect performance of the spatial
resolution on ILD-TPC by 2010 test beam results.  
Case (a) in FIG.~\ref{fig:extrpol} is estimated by Magboltz for B = 3.5 T
with the same configuration as our test beam. 
Case (b)  in FIG.~\ref{fig:extrpol} is not based on
any real detector, but it can be realized if we take larger gap between 
GEMs and the stronger electric field in the gap.
The expect performance of the spatial resolution based on our test beam result
can be expected to satisfy the requirement ($\sigma \leq 100 \ \mu m$) of 
ILD-TPC for almost drift region.

\section{Local field distortion}
\begin{figure}[htbp]
\centering
\subfigure[2010 test beam]
{\includegraphics[width=8.15cm, bb=0 0 426 246]{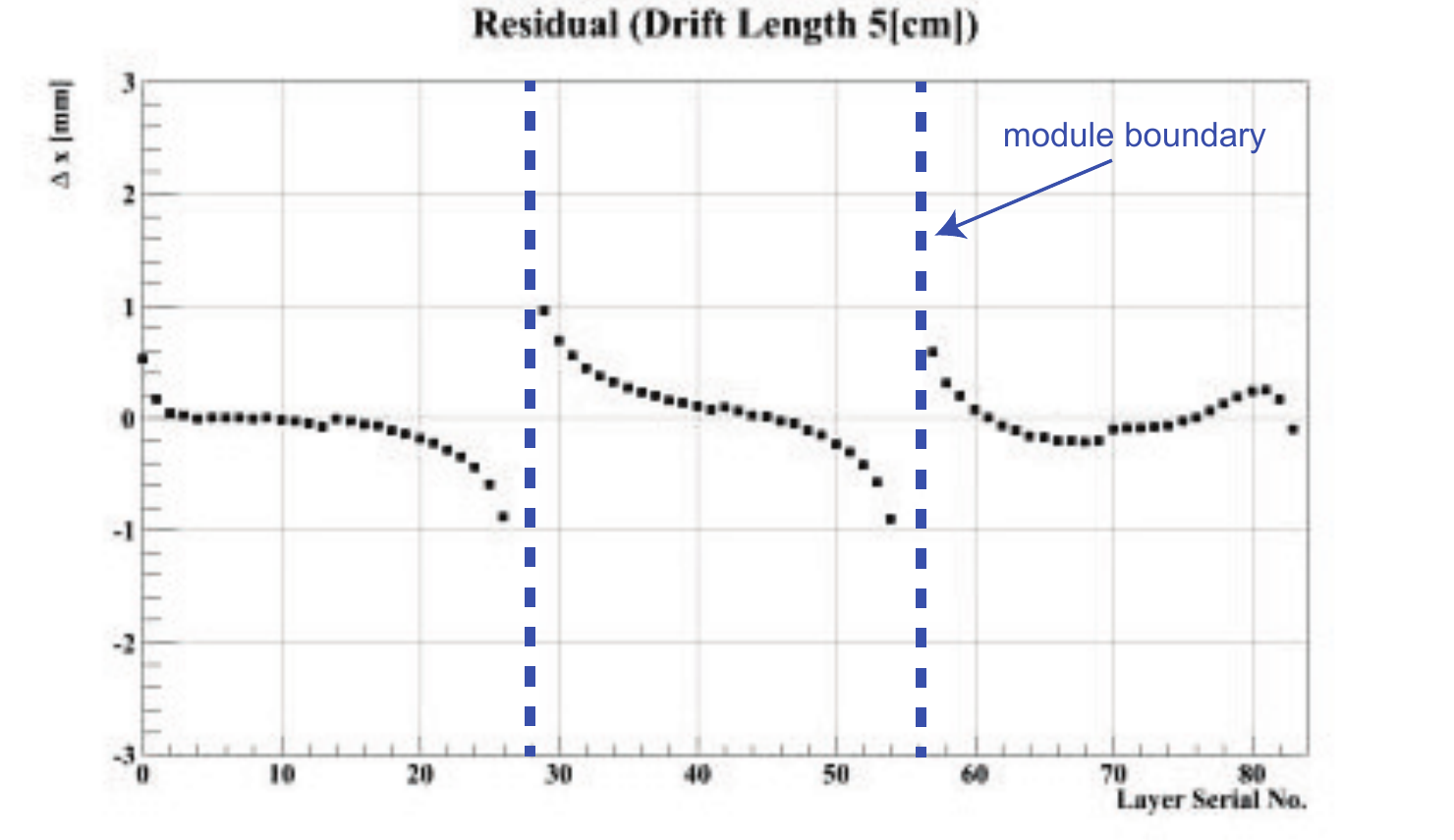}}
\subfigure[2012 test beam]
{\includegraphics[width=8.15cm, bb=0 0 373 214]{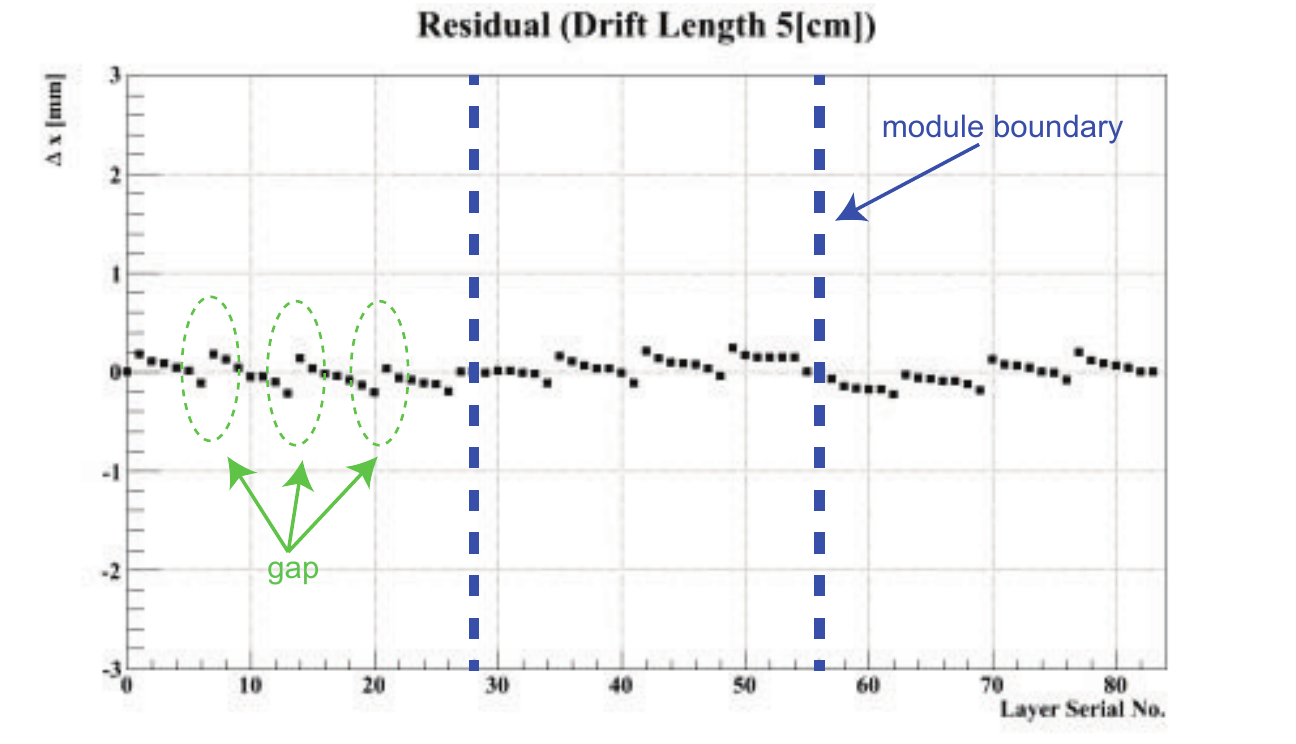}}
\caption{The residual distribution of three modules}
\label{fig:distortion}
\end{figure}

As mentioned in the previous section, 2010 test beam data has a big distortion on all modules in FIG.~\ref{fig:distortion}~(a).
The distortion has occurred around the module boundary. The cause of distortion
 is
wrong setting of HV on field shaper. Since the distortion makes the worse of the spatial
resolution, the decrease of the distortion is important. In 2012 test beam,
GEM module is increased the partitions of electrode (4 partitions) and optimized
HV of field shaper to suppress the distortion. FIG.~\ref{fig:distortion}~(b) shows there is a smaller distortion on 
the module boundary than 2010, but the small distortion is appeared at the electrode
boundary in GEM. The size of electrode gap of 2012 GEM module is 1 mm but the size of 2010 is 
200~$\mu$m. Our simulation study found this wide gap has disturbed 
the field uniformity 
around the gap. The good field uniformity needs more narrow gap.  
We improved the 
new GEM module to suppress the distortion for next test beam. The improved GEM 
has the boundary gaps on the front side only and changes gap size to 500 $\mu$m 
from 1 mm. Since it is important to measure the distortion by other method before 
the test beam, we make the UV laser system to measure it. The test plans with
UV laser system measure the distortion of improved GEM and compare to simulation
and measure the distortion of GEM with wire gate.

\section{Effects of positive ion and gate devices}
Positive ion effect is one of main  problems of TPC to get the excellent spatial 
resolution. Positive ion generated by ionization and gas amplification 
move to the cathode and
disturb the field uniformity in the drift region. TPC with MWPC has to be equipped 
the gate device because all ions goes back. Since there are a lot of ions in TPC at the 
ILC, we have investigated the ion effect on TPC with MPGD(GEM)~\cite{Gros-ECFA}. 

\begin{figure}[htbp]
\centering
 \subfigure[primary ion (associated with track)]
  {\includegraphics[width=10cm, bb=0 0 495 326]{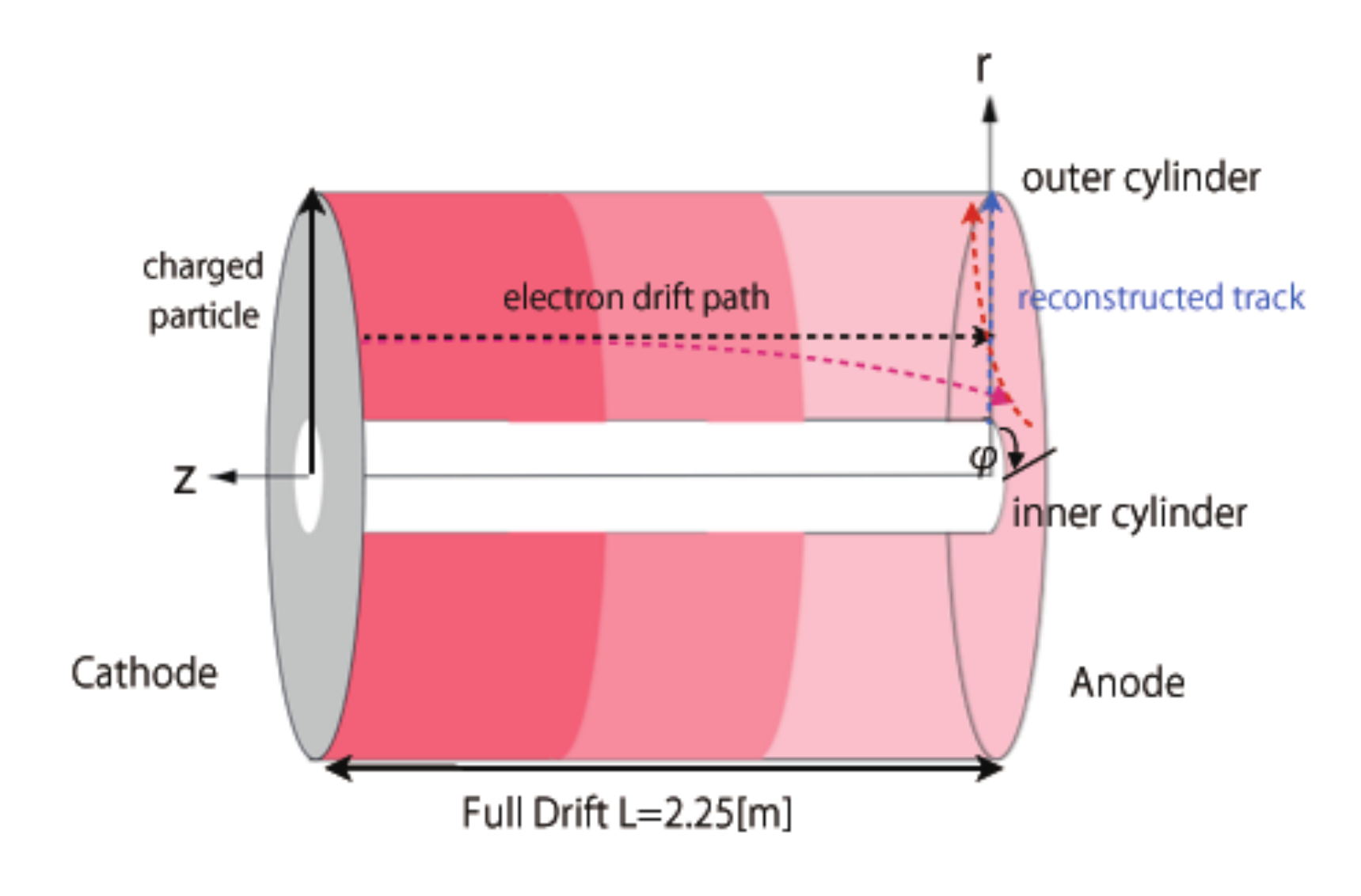}}
\subfigure[secondary ion (associated with GEM)]
 {\includegraphics[width=10cm, bb=0 0 511 310]{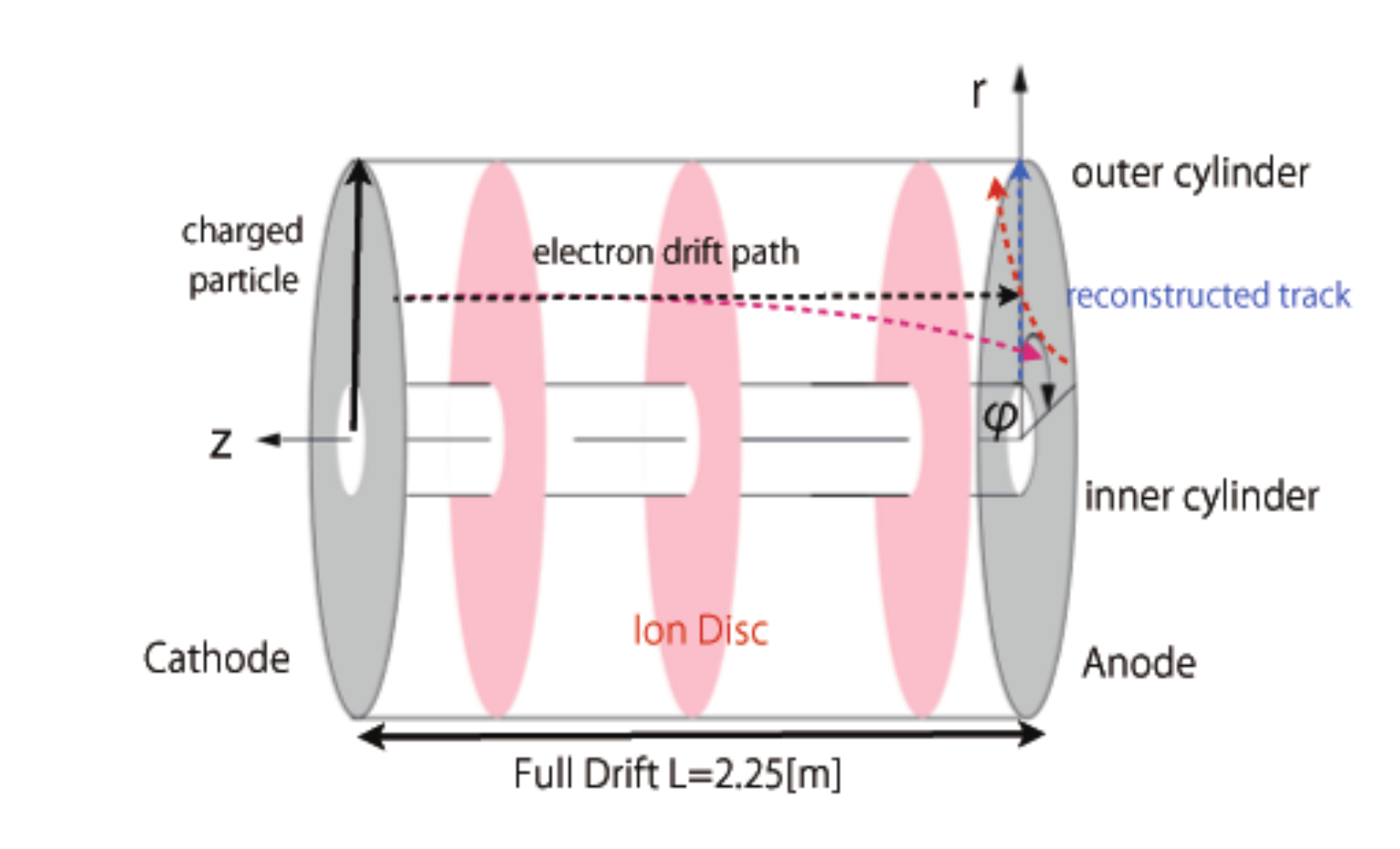}}
\caption{Positive ion effect on ILD-TPC}
\label{fig:ioneffect}
\end{figure}

The method of the estimation of the ion effect is, 
\begin{enumerate}
\item make the ion density distribution with the signal and background simulation
\item estimate the field distortion by solving the Poisson equation with proper
boundary conditions
\item evaluate the distortion of drift electron trajectory from the field distortion
by the Langevin equation
\end{enumerate}
There are two type of ion, the primary ion and the secondary ion, which have 
different origin and effects on the positive ion effect. The primary ion comes from
the ionization by the charged particle and moves to the cathode. The frequency of a 
beam train in ILC is 5 Hz~\cite{ILC-DBD} and the drift region existed the three primary ion
cylinder by three a beam train at same time. FIG.~\ref{fig:ioneffect}~(a) illustrates the
primary ion effect on ILD-TPC.
The secondary ion comes from the gas amplification and the secondary ion density 
is calculated with (the number of primary ion) $\times$\ (ratio of the entering of the primary ion
to gas amplification) $\times$\ (gas amplification factor).  In our estimation, the density of
the secondary ion is about 2 to 9 times bigger than the density of the primary ion.
The distribution of the secondary ion is different from the primary ion as FIG.~\ref{fig:ioneffect}~(b).
The secondary ion by a beam train in ILC forms a disk of thickness 4 mm and moves to the
cathode. The drift region has three ion disk at same time on ILD-TPC.  
Using the above simulation results, the distortion of the track is estimated for the both case.
TABLE~\ref{tab:iongate} shows the distortion of the reconstructed track by the positive ion. 
\begin{table}[htbp]
\begin{center}
\caption{Distortion of track by positive ion}
\label{tab:iongate}
\begin{tabular}{|c|c|c|}
\hline
 & {\bf without} gating device & {\bf with} gating device \\  
 \hline
 Primary ion & 8.5 $\mu$m & 8.5 $\mu$m \\
 \hline
 Secondary ion & 60 $\mu$m & 0.01 $\mu$m \\
 \hline
 sum & 70 $\mu$m & 8.5 $\mu$m \\
 \hline
 \end{tabular}
 \end{center}
 \end{table}
 The primary ion doesn't affect the distortion of the track. And the gating device can't 
 suppress the distortion by the primary ion. The secondary ion make much effort to the
 distortion. The gating device can suppress to the distortion. Therefore ILD-TPC has to 
 need to equip the gating device which suppress to the distortion by the secondary ion. 
 
ILD-TPC requires the performance of the ion gate devices to (1) ion feedback must be
smaller less than 10$^{-3}$, (2) gate can open for 1~msec and close following 199~msec,
(3) drift of ion can be less than 1~cm. There are 3 candidates  of the gate devices, wire, 
GEM, and micromesh,  suitable for ILD-TPC. 

\begin{figure}[htbp]
\centering
\subfigure[Picture of wire gate module]
 {\includegraphics[width=4.5cm, bb= 0 0 640 480]{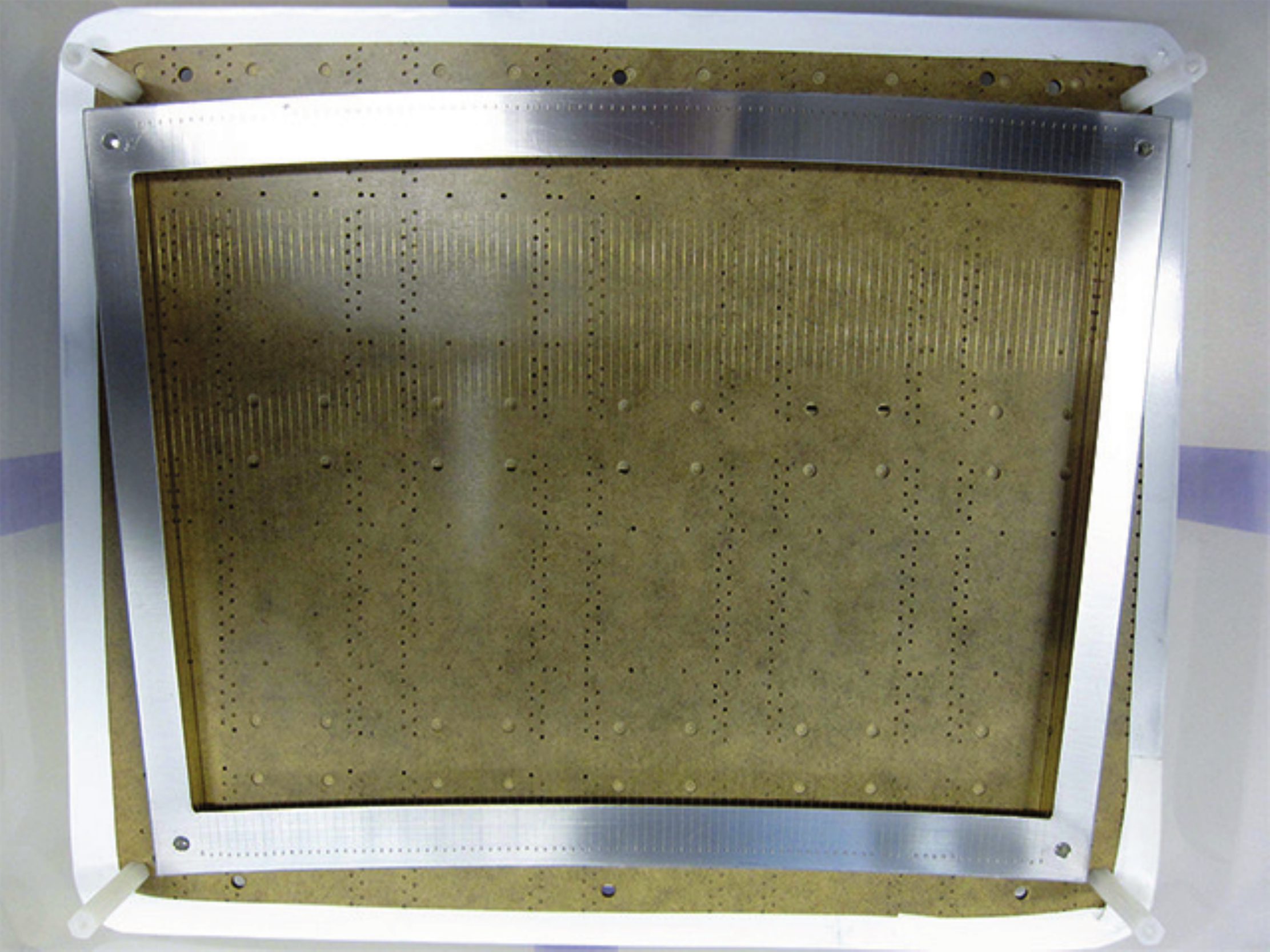}}
\subfigure[charge distribution of Fe55]
  {\includegraphics[width=8cm, bb= 0 0 696 472]{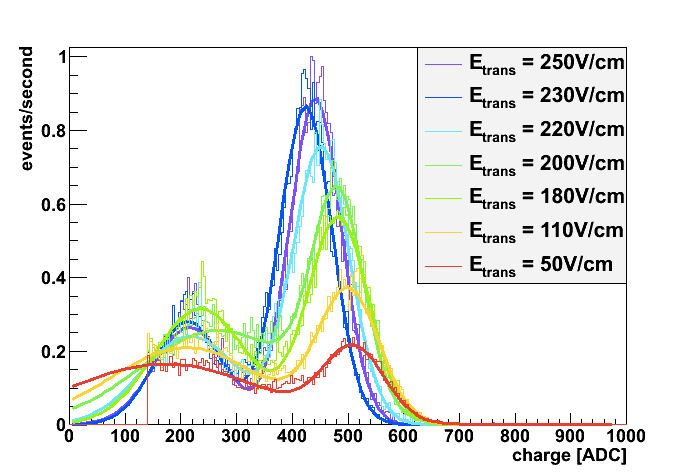}}
\caption{wire gate module and it's  charge distribution}
\label{fig:wiregate}
\end{figure}

The wire gate is used TPC with MWPC.
The effect of an open wire gate on the
electron trajectories should be negligible, even in a high magnetic field. 
Since the existing  wires deteriorate electric field around a wire, some $E \times B$\
effect may happen at around a wire. Building a wire 
gate independently from MPGD read out modules would require extra structure,
and would be difficult to achieve without introducing problematic dead region.
ILD-TPC also requires a wire gate to be included within the MPGD read out modules.
 There are two different configurations that can be considered, transverse wires and
 radial wires. The radial wire system is preferable for the small modules of ILD-TPC.
The support structure would be mainly transverse to the tracks, so that the dead 
region would not interfere the tracking coverage. Our group made a radial wire gate
module and tested it with Fe55. FIG.~\ref{fig:wiregate}~(a) is picture of our wire
gate module.
A wire of 30 $\mu$m in diameter is put on this gate every 2 mm and spot
welded on stainless steel frame (up and down). This only applies a single potential regime for
the closed gate configuration. Using this gate, we measured the Fe55 spectra 
for different HV configuration with fixed drift field (230 V/cm). 
FIG.~\ref{fig:wiregate}~(b) is the charge distribution of Fe55. This figure shows 
the decrease of the transfer (the region between gate and GEM) field lead to the decrease of the transmission efficiency.

\begin{figure}[htbp]
\centering
 \includegraphics[width=4.5cm, bb=0 0 992 755]{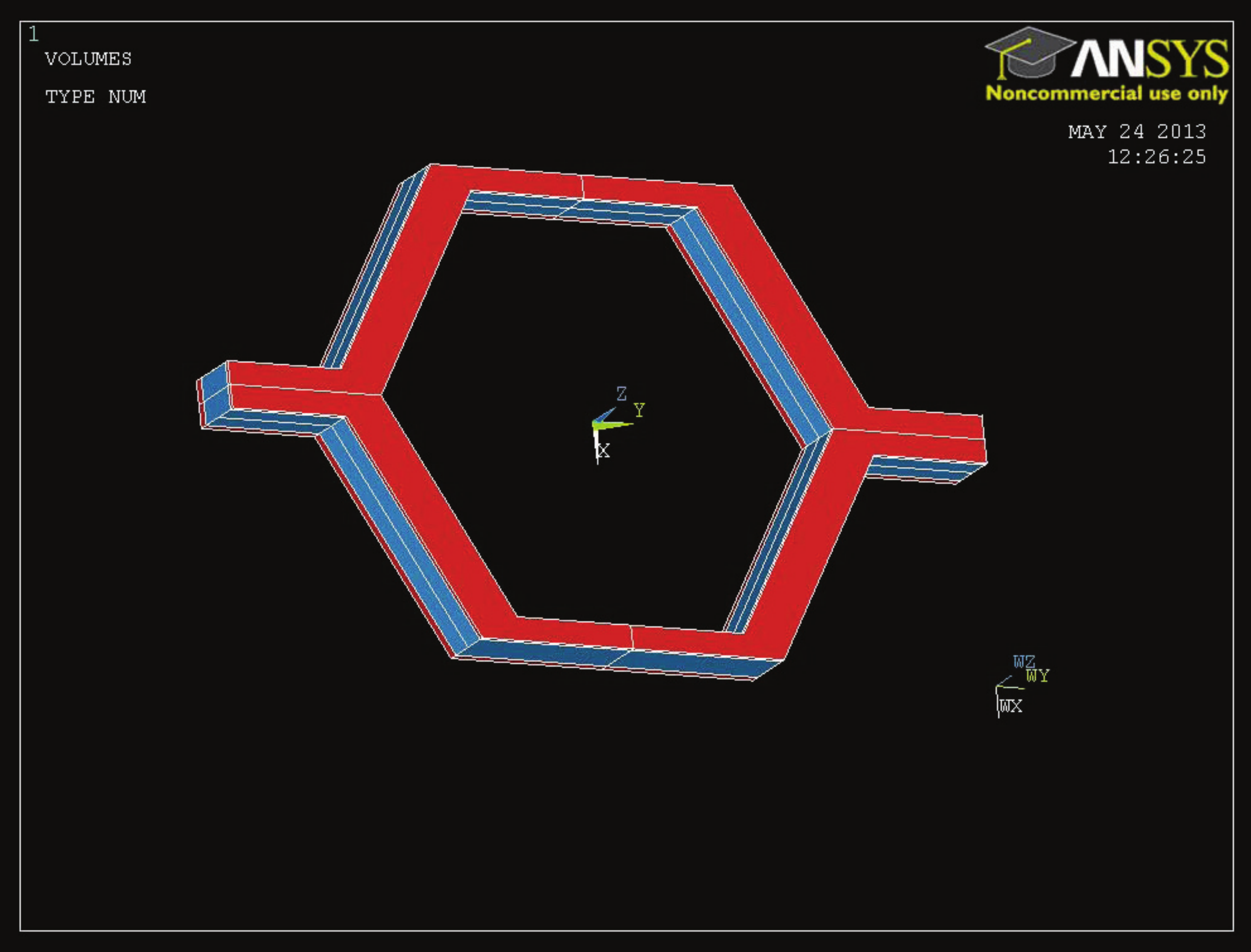}
\caption{GEM geometry simulated. Only one sector is used for field calculations, using symmetries.}
\label{fig:GEMgate}
\end{figure}

GEM gate is a good candidate for TPC with GEM because it is almost same 
frame and can be stacked on GEM module. F.Sauli suggested to use GEM as
the ion gate devices in TPC~\cite{GEMgate}. The key issue of GEM gate is 
the electron transmission.　We made a thin GEM gate of 14 $\mu$m thick with 
90 $\mu$m holes and 140 $\mu$m pitch and measured the electron 
transparency~\cite{gateTEST}.
This GEM offers 37\% geometrical aperture. This measurement
has done with a 1 T magnetic field and T2K gas (${\rm Ar:CF_{4}:iC_{4}H_{10} = 
95:3:2}$).  This measurement shows a maximum transparency to be 
 below 40\% around 10 V. The simulation can describe this results and the 
 increase of aperture size has  been a key parameter in a magnetic field.
Getting the satisfied transparency, we simulate a GEM geometry with a maximized aperture
as a hexagonal honeycombed structure in FIG.~\ref{fig:GEMgate}~\cite{gateTEST}.
This structure offers an 81 \%  aperture, but it is difficult to build this structure with thin devices.
As for the simulation, the electric field was calculated in ANSYS on one sector of the GEM, using
symmetric boundaries and the Garfield++ simulation uses these symmetries to extrapolate 
the field to the whole space.
 
\begin{figure}[htbp]
\centering
\subfigure[Electron transmission for different transfer field]
 {\includegraphics[width=8.15cm, bb= 0 0 1916 959]{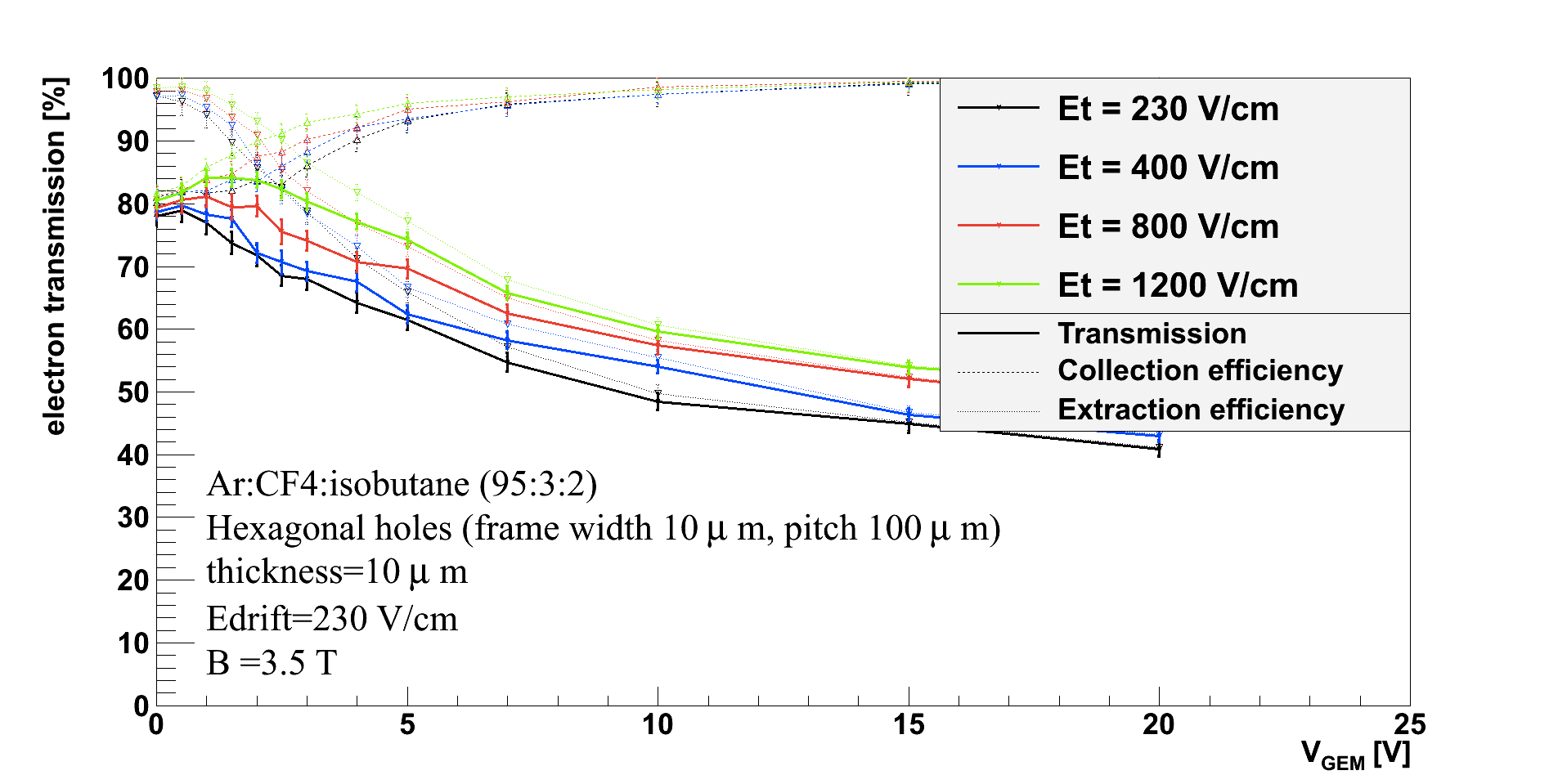}}
\subfigure[Electron transmission for different GEM thickness]
 {\includegraphics[width=8.15cm, bb= 0 0 1916 959]{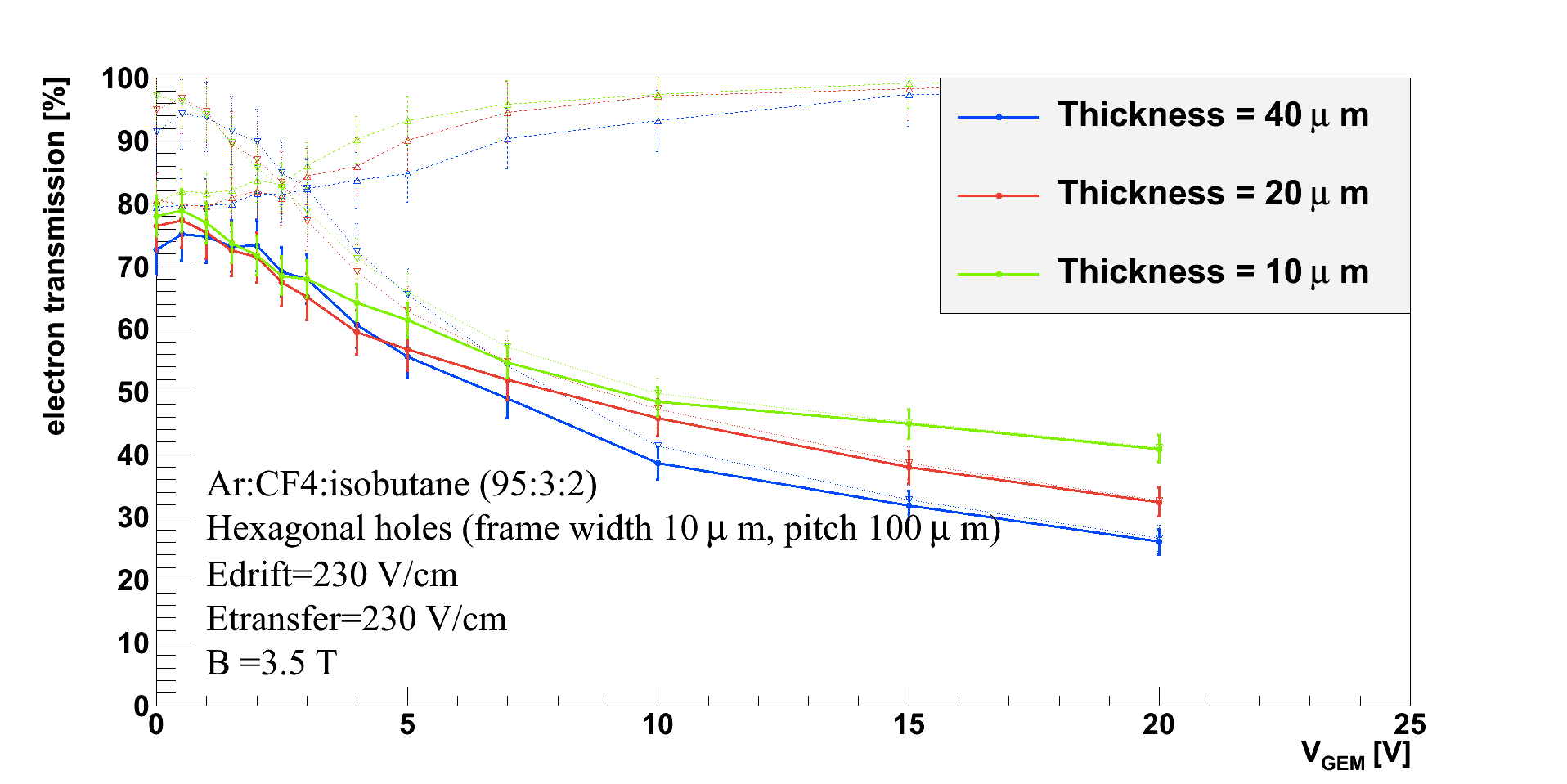}} 
\caption{Results of simulation with large aperture GEM in a 3.5T magnetic field.}
\label{fig:GEMgatesim}
\end{figure}

We measured the simulated electron transmission efficiency for different transfer field and
different thickness of the GEM~\cite{gateTEST}. FIG.~\ref{fig:GEMgatesim} shows the influence
of these parameters is as expected with 3.5T magnetic field.
These figures show the maximum transmission efficiency is reached above 70\% in the special
conditions (very large aperture, very thin structure, a higher transfer field).  A transmission 
efficiency close to the geometrical aperture of the GEM can always be achieved for low
voltage. Due to the large $\omega\tau$\ of T2K gas in the large magnetic field, the
transfer field and thickness have little influence on the transparency.
We will challenge to product a GEM gate with suitable mechanical properties of ILD-TPC.
This might be possible with large holes and relatively thick GEMs.

\section{Cooling for electronics}
ILD-TPC group is investigating to introduce the 2-phase CO$_{2}$\ cooling system as the cooling
for the read out electronics. The advantages of the 2-phase CO$_{2}$\ cooling are,
(1) the latent heat of the liquid CO$_{2}$\ is very big with 300 J/g, (2) coverage of cooling is
down to -40 degree from room temperature, (3) using the latent heat for cooling can suppress 
a raise in cooling temperature by a temperature rise of the refrigerant, (4) using of a 
high-pressure refrigerant and small liquid viscosity can use thin pipe for the cooling system
 and deduct the material budget, (5) it is free from worry about the short circuit of the detector even if
 there is a leak of the refrigerant.
 
KEK CO$_{2}$\ group~\cite{KEK-CO2} took a demonstration experiment of simplified
2-phase CO$_{2}$\ cooling system in 2011 and 2012 and confirmed the ability of cooling
power. ILD-TPC group made the small 2-phase CO$_{2}$\ cooling system with 2-phase 
CO$_{2}$\ circulation pump in cooperation with NIKHEF in 2012. This system put on
ILD-TPC large prototype test facility in DESY and use the heat control for electronics.
We also study of cooling system for S-ALTRO16 read out electronics with thermal simulation.

\section{Summary and future plan}
ILD-TPC Asia group has been studying the high precision TPC with GEM 
as a central tracker of ILC-ILD as a member of ILD-TPC collaboration. 
We took the test beam using the large 
prototype TPC in DESY at 2010 and 2012. The both results are consistent with
each other taking into account of the real charge loss at 2012. 
Analytic formula for TPC with GEM has been evolved and explains the
results of test beam data. Using the analytic formula, the prospective 
performance estimated from the test beam data can be expected to satisfy ILD-TPC performance.
We will take the test beam with $B=3.5$\ T magnet with a new read out 
electronics and 2-phase CO$_{2}$\ cooling system.

The GEM used by test beam of 2010 and 2012 had occurred the distortion problem.
Since the distortion makes the spatial resolution to be worse, 
we design the new GEM module which suppress the distortion.
And our laser test system can measure it before the test beam.

 Development of the gate device is a key issue for ILD-TPC. The 
requirement of spatial resolution requires to suppress the secondary ion 
effect and to have the gate device. The GEM gate is a strong candidate for
ILD-TPC with GEM. The simulation study of GEM gate shows the geometrical
aperture ($>$\ 70\%) is the key parameter in high B field. The main problem
for GEM gate is building such large aperture GEM. We will make a large 
aperture GEM gate and measure the electron transparency.

The study of 2-phase CO$_{2}$\ cooling system has just started. We build
the colling system to check the cooling ability and this system will be
used test beam at DESY. The design of cooling system is an important
R\&D items. We are studying the cooling system for the electronics in
cooperation with ILD-TPC collaboration.

\begin{acknowledgments}
This work was supported by the grant-in-aid specially promoted research
No.23000002 of the Japan Society of Promotion of Science.
\end{acknowledgments}

\end{document}